Fitting analysis provides further evidence for eradication of hiv/aids infection under combined liposome drug delivery treatment

Evgeni E. Gabev[1]*, Evgeni B. Gabev[2], Evgeni E. Gabev Jr.[1], Masha V. Bogoeva[3]

[1]Institute of Experimental Pathology and Parasitology, Bulgarian Academy of Sciences Sofia 1113, Bulgaria

[2]Department of Infectious Diseases, Epidemiology, Parasitology and Tropical Medicine, Medical University, Sofia 1431, Bulgaria.

[3]Institute of Zoology, Bulgarian Academy of Sciences, Sofia 1000, Bulgaria.

*To whom correspondence should be addressed. E-mail: egabev@bas.bg

**It is now evident that the commonly accepted strategy for treatment of HIV/AIDS by highly active antiretroviral therapy (HAART) will not lead to eradication of HIV in a reasonable time. This is straightforward from the typical exponential viral load decay upon treatment revealing initial considerable but incomplete reduction of plasma HIV RNA with subsequent low level HIV persistence even in patients on effective anti-retroviral therapy. Here we show that the viral load follows a simple zero trend linear regression line under different treatment approach recently proposed by us. This unambiguously indicates a whole body HIV eradication in reasonable time.**

Recently we proposed and evaluated a new approach for treatment of HIV/AIDS (*1*). In contrast to HAART it is based on simultaneous destroying the virus (both in the blood stream and its reservoirs) and rendering the non infected target cells refractory to HIV attack as well. The later was accomplished by blocking the cell's polyphosphoinositide transmembrane signaling system and, in turn, the generated second messengers, calcium release and the triggering of protein kinase C (PKC). The crucial role of the polyphosphoinositide pathway in HIV life cycle is well documented in the literature (*2-7*). We implemented this approach in practice using our liposome anti HIV/AIDS preparation FTL/AZT/PEBA containing lithium as a specific non-competitive blocker of the polyphosphoinositide pathway (*8, 9*) and 3'-azido-3'deoxythymidine (AZT, azidothymidine) as an antiretroviral agent, both of them obligatory encapsulated in liposomes. The results obtained by extended evaluation of our preparation on HIV infected cell cultures, experimental animals and AIDS suffering subjects showed that it is not toxic and such approach may contribute considerably to successful AIDS therapy (*1*).

The primary goal of the present work is to evaluate and interpret the best fit line for our experimental data of viral load measured in AIDS patients under treatment with FTL/AZT/PEBA aiming to provide additional evidence for its therapeutic abilities, advantages and mode of action as well as to use the graph as a tool for making prognosis of therapy outcome and to estimate the time necessary to reach the end point of the medication period. Figure 1 illustrates that upon treatment with FTL/AZT PEBA the plasma HIV-1 RNA follows a simple downward (negative slope) linear regression line instead of the exponential one which is typical when non liposome encapsulated (free) antiretrovirals, e. g. HAART are used and in contrast to ours does not tend to zero viral load (*10, 11, 12, 13*). Taking into account our previous findings (*1*) and the continual downstream nature of the straight line (Fig. 1) we came to a conclusion that a whole body eradication of HIV could be

3obtained in reasonable time. Most possibly this could be explained by the advantageous features of the liposome drug delivery (*14, 15, 16*) and lithium dual action on both HIV-to-host cells signalling (*2-7*) and cytoplasmic HIV RT DNA conformational state(*1*). Briefly, some of the liposomes carrying the active substances (lithium and AZT in this particular case) are phagocytized by the macrophages both in the blood circulation and in the HIV sanctuary organs with subsequent lysis and sustained release of the above components. Other simply adhere to the cell outer membrane surface whereupon the drug molecules will diffuse through the liposome lipid bilayer and into the cell. Similar role plays the liposome-to-cell fusion (*16*). The above mechanisms of liposome drug delivery are of particular importance for creating intracellular constant optimal therapeutic concentrations of AZT and lithium both into the phagocytic cells as well as in non phagocytic HIV target cells e. g. CD4+ T lymphocytes. Upon contact with HIV these intrinsically quiescent cells become activated and are forced to proliferate via specific receptor signaling mechanism which is promoted by the polyphosphoinositide pathway (*3, 4, 6*). In result productive and/or latent HIV infection is developed. Liposome delivered lithium stops such an activation process thus rendering the target non infected CD4+ T cells refractory to HIV attack. This is achieved by inhibition the key enzymes of the phosphoinositide pathway the inositol monophosphatase (IMP) and inositol polyphosphate 1-phosphmonoesterase (*7, 9*). Besides, the intracellular liposome delivered lithium ions may add positive charge to the negative phosphate groups of the cytoplasmic HIV RT DNA with consequent conformational changes and inactivation of the molecule (*1*). There are both experimental and computer simulation evidence that such interactions may take place (*17*). Apparently, further integration of the cytoplasmic HIV DNA into the cell genomic one will thus be avoided. The above mechanism provides hope that one of the most stubborn HIV reservoirs the latently infected CD4+ T cells (*10*) might be effectively affected thus contributes to whole body HIV eradication. In the same time HIV DNA production is successfully blocked by AZT especially being liposome delivered into the target cells. Recently a subset of human natural killer (NK) cells that express CD4 and HIV co receptors CCR5 and CXCR4 have been identified and proved susceptible to HIV-1 infection in a CD4-dependent manner (*18*). The presented results provide strong evidence that such NK cells remain persistently infected with HIV-1 even in patients receiving HAART for 1-2 years. Apparently the above described mechanism of action of FTL/AZT/PEBA may also contribute to elimination of such non T cell HIV reservoirs.

As illustrated (Fig. 1) the prognosis of the therapy outcome might be considered optimistic in terms of reaching complete cure as long as the viral load progression follows a downward linear shape with zero trend. Stable deviations from this regression model may indicate for instance either possible development of drug resistance or HIV rebound due to replenishment from ineradicable reservoir(s) with respective failure of the treatment. We investigated the theoretical time for zero viral load and obtained a reasonable on treatment period of about 50 weeks (Fig. 1, A to C). Since there is not yet available an assay with detection limit bellow 25-50 plasma HIV-1 RNA copies per milliliter and moreover that our PCR limit is 500 copies/ml we also calculated the time necessary to reach a viral load which is four and five orders of magnitude bellow the mathematical zero. Nevertheless we obtained a reasonable on treatment periods depending on the viral load (Fig. 1) which are far bellow than the estimations based on HAART (over 60 years in average) (*10*).

In conclusion we show that the linear zero trend viral load decay obtained for the first time under our HIV/AIDS treatment approach (*1*) suggests that a whole body eradication of the virus is achievable in reasonable time. We consider the following main points in support of that. First, the ability of FTL/AZT/PEBA for simultaneous and complete knock out of HIV from both reproductively infected and reservoir cells regardless of their type and organ localization based on the liposome drug delivery mechanisms described above. Second, rendering of the non infected target cells refractory to HIV attack by blocking their polyphosphoinositide transmembrane signalling system thus preventing infection expansion and improving the immune system status. And last but not least, creating conformational changes in de novo synthesized cytoplasmic HIV RT DNA with consequent inactivation of the molecule thus stopping both further reproductive and/or latent HIV infection. The graph may also serve as an useful tool for early prediction of drug resistance and managing the therapy.

Figure legend.

Figure 1 The estimated best fit line (linear regression) of the viral load (VL) decay obtained under treatment with FTL/AZT/PEBA. (**A**) Cumulative curve for 8 AIDS patients P = 0.041, $r^2$ = 0.25. The calculated time for obtaining zero VL is 51 weeks and 53 and 83 weeks for four and five orders of magnitude bellow zero respectively. (**B**) Representative example for patient (HIS) whose VL reached the detection limit of our PCR (500 copies/ml HIV-1 RNA, indicated by horizontal dashed line) within 46 weeks of treatment (P = 0.0041, $r^2$ = 0.8973). We calculated that 52 weeks are necessary so that the patient's VL to reach the mathematical zero and respectively 71 and 244 weeks for 3 and 4 orders of magnitude bellow zero. (**C**) Representative example for patient (MAS) whose VL sharply decreased but did not reach the detection limit of our assay within 31 weeks of treatment. We calculated that 38 weeks are necessary to obtain zero VL and respectively 46 and 116 weeks for VL of four and five orders bellow the mathematical zero. We used Curve Expert version 1.36 (Hyams Development, USA) for preliminary automatic fit with 36 models and interpolation the viral load with time. We used Prism version 2.01 (*20*) for further detailed analyzing of the best fit linear regression line and estimation the goodness of fit parameters and calculation the theoretical end point time necessary to obtain different viral loads. We also used Prism to construct the graphs. Amplicor HIV-1 Monitor test (Roche Diagnostic Systems, USA) was used for measuring of HIV-1 RNA copies.

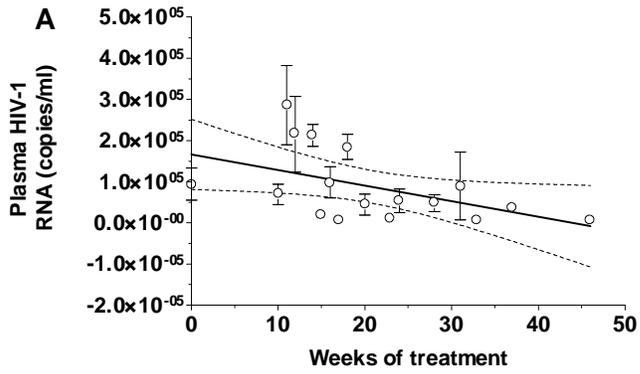

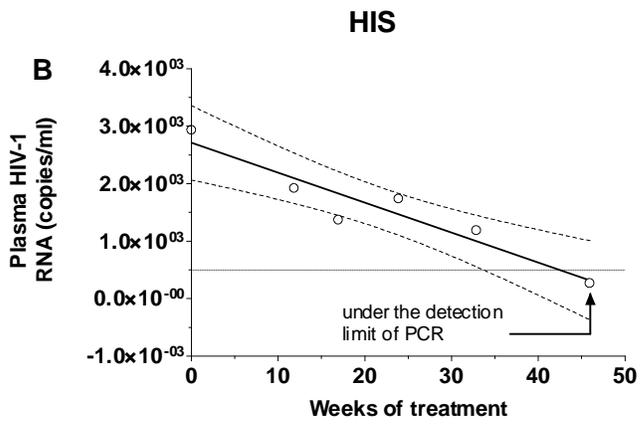

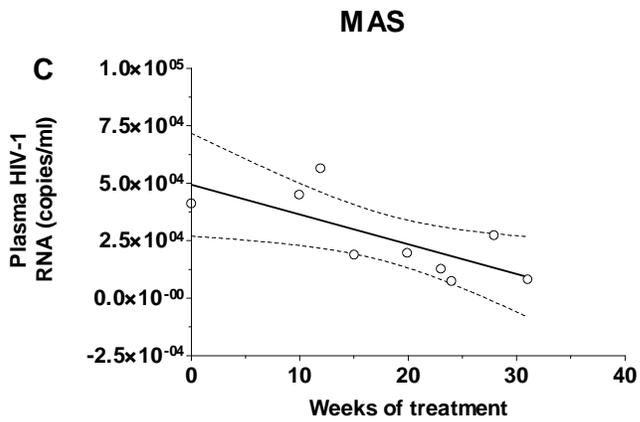